\documentclass[article,aps,nofootinbib,showpacs,amsfonts,epsf]{revtex4}

\input{epsf.tex}

\newcommand{\E}{{\cal{E}}}

\newcommand{\s}{\sigma}

\renewcommand{\a}{\alpha}
\renewcommand{\k}{\kappa}
\newcommand{\dfrac}[2]{\displaystyle\frac{#1}{#2}}
\newcommand{\be}{\begin{equation}}
\newcommand{\ee}{\end{equation}}
\newcommand{\bea}{\begin{eqnarray}}
\newcommand{\eea}{\end{eqnarray}}
\newcommand{\ba}{\begin{array}}
\newcommand{\ea}{\end{array}}
\def\J#1#2#3#4{{#1} {\bf #2}, #3 (#4)}

\def\PR{Phys. Rev.}
\def\PRL{Phys. Rev. Lett.}
\def\PTP{Prog. Theor. Phys.}
\def\APN{Ann. Phys. (NY)}

\def\JMP{J. Math. Phys.}

\def\CQG{Class. Quantum Grav.}

\def\PLA{Phys. Lett. A}

\begin{document}
\draft

\title{On the physical interpretation of the $\delta=2$ Tomimatsu-Sato solution}

\author{V.~S.~Manko}
\address{Departamento de F\'isica, Centro de Investigaci\'on y de
Estudios Avanzados del IPN, A.P. 14-740, 07000 M\'exico D.F.,
Mexico}

\begin{abstract}
The physical properties of the Tomimatsu-Sato $\delta=2$ spacetime are analyzed, with emphasis on the issues of the negative mass distribution in this spacetime and the origin of a massless ring singularity which are treated with the aid of an equatorially asymmetric two-body configuration arising within the framework of the analytically extended double-Kerr solution. As a by-product of this analysis it is proved analytically that the Kerr spacetime with negative mass always has a massless naked ring singularity off the symmetry axis accompanied by a region with closed timelike curves, and it is also pointed out that the Boyer-Lindquist coordinates in that case should be introduced in a different manner than in the case of the Kerr solution with positive mass.  \end{abstract}


\maketitle


\section{Introduction}
The famous $\delta=2$ Tomimatsu-Sato (TS2) solution \cite{TSa1,TSa2} discovered in 1972 within the framework of Ernst's complex potential formulation of the stationary axially symmetric gravitational field problem \cite{Ern} describes the exterior field of a spinning mass different from that of a Kerr black hole \cite{Ker}, and it has been discussed extensively in the literature. In the paper of Gibbons and Russell-Clark \cite{GRu} it was suggested that the massive source of TS2 solution could be a specific rotating star; at the same time, the relevance of the solution to black-hole physics was discarded in that paper because of the presence in TS2 spacetime of a region with closed timelike curves (CTC) containing a naked ring singularity outside the symmetry axis, and also because the hypersurface $x=1$ was found not to be an event horizon of the solution.

Several years ago, however, the interest in the global properties of TS2 spacetime in the black-hole context was revived by Kodama and Hikida \cite{KHi} who pointed out that the poles $x=1$, $y=\pm1$ of the TS2 solution were in fact two degenerate Killing horizons, whence they concluded in particular that the TS2 solution could provide ``a new possibility for the final states of gravitational collapse'', the latter idea being recently reiterated in \cite{BYo,GLS}. In this respect it is probably worth reminding that essentially all non-trivial members of the double-Kerr (DK) solution of Kramer and Neugebauer \cite{KNe}, to which the TS2 spacetime does belong, independently of their ``pathologies'' possess two Killing horizons (see, e.g., Ref.~10), p.~348), and, besides, that the binary black-hole configurations constitute only a 5-parameter subfamily of the general 8-parameter DK solution, to which the TS2 spacetime does not belong. Consequently, in our opinion, the most important physical aspects of the TS2 solution that are likely to be clarified first of all are the nature of the massless ring singularity and origin of the CTC region because providing a reliable explanation for these phenomena would also shed additional light on many other exact solutions with similar properties. In the present paper we shall give a convincing demonstration that there exists a direct relation between the aforementioned two phenomena and the negative mass of the TS2 solution. This will be done, on the one hand, by obtaining analytical formulas for the mass distribution in the TS2 spacetime and, on the other hand, by considering the evolution of a one-parameter generalization of the TS2 solution, representing a specific non-linear superposition of two Kerr-NUT sources, which will enable us to trace the emergence of the ring singularity on the stationary limit surface (SLS) of the constituent with negative mass.

The paper is organized as follows. In Sec.~2 we give a brief review of the TS2 solution and some of its known properties. Here we observe in particular that the mass quadrupole moment of TS spacetimes has never been given correctly in the literature. In Sec.~3 the mass and angular momentum distributions of the TS2 solution are considered, establishing in this manner the presence of a negative Komar mass introduced by the non-zero NUT parameters of the corresponding DK solution. A one-parameter generalization of the TS2 solution which leaves the total mass and total angular momentum of the latter unchanged is constructed and analyzed in Sec.~4. This new binary configuration permits us, by varying the parameter $z_0$ from the larger to smaller values, to show that the appearance of the ring singularity on the SLS of the lower Kerr-NUT constituent takes place when the mass of this constituent changes its sign from plus to minus. The TS2 solution with negative total mass is considered in Sec.~5, and in this case two regions with CTCs appear, one touching the SLS outside the symmetry axis and the other not. Then the ring singularity arising only in the former region is subsequently given a generic interpretation as a locus of points at which the SLS touches the boundary of a CTC region. To support this definition of a ring singularity in the vacuum spacetimes we consider the Kerr solution as an additional example. Finally, Sec.~6 is devoted to discussion of the results obtained and conclusions.

\section{The TS2 metric and its known properties}

The TS2 metric in the Ernst-Perj\'es representation \cite{Ern1,Per} is defined by the line element
\be ds^2=\kappa^2f^{-1}\left[e^{2\gamma}(x^2-y^2) \left(\frac{d
x^2}{x^2-1}+\frac{d y^2}{1-y^2}\right) +(x^2-1)(1-y^2)d\varphi^2\right] -f(d t-\omega d\varphi)^2, \label{Papa} \ee
where the functions $f$, $\gamma$ and $\omega$ are given by the formulas
\bea f&=&\frac{\mu^2-(x^2-1)(1-y^2)\s^2}{\mu^2+\mu\nu-(1-y^2) [(x^2-1)\s^2-\s\tau]}, \nonumber\\ e^{2\gamma}&=&\frac{\mu^2-(x^2-1)(1-y^2)\s^2}{p^4(x^2-y^2)^4}, \nonumber\\ \omega&=& -\frac{\k(1-y^2)[(x^2-1)\s\nu+\mu\tau]}{\mu^2-(x^2-1)(1-y^2)\s^2}, \label{mf_pol} \eea and the polynomials $\mu$, $\s$, $\nu$, $\tau$ have the form \bea &&\mu=p^2(x^2-1)^2+q^2(1-y^2)^2, \quad \s=2pq(x^2-y^2),  \nonumber\\ &&\nu=4x(px^2+2x+p), \quad \tau=-4qp^{-1}(1-y^2)(px+1), \label{pol4} \eea
the real constants $p$ and $q$ being subjected to the constraint $p^2+q^2=1$. Prolate spheroidal coordinates $x$, $y$ are related to the Weyl-Papapetrou cylindrical coordinates $\rho,z$ by the equations
\be
x=\frac{1}{2\k}(r_++r_-), \quad y=\frac{1}{2\k}(r_+-r_-), \quad r_\pm=\sqrt{\rho^2+(z\pm\k)^2}, \label{xy} \ee
where $\k$ is the positive real constant in the line element (\ref{Papa}). The inverse transformation is
\be \rho=\kappa\sqrt{(x^2-1)(1-y^2)}, \quad z=\kappa xy. \label{xy_inv} \ee

The total mass $M$ and total angular momentum $J$ of TS2 solution are given by the expressions \cite{TSa1}
\be
M=\frac{2\kappa}{p}, \quad J=M^2q, \label{MJ} \ee
and, whereas the whole mass is contained in the central rod $x=1$, $y^2<1$, the angular momentum is distributed between the rod and the two poles $x=1$, $y=\pm1$ in the following way \cite{GLS}:
\be
J=J_{\rm rod}+2J_{\rm pole}, \quad J_{\rm rod}=\frac{2\kappa^2(2-p)(1+p)}{p^2q}, \quad J_{\rm pole}=-\frac{\kappa^2(1+p)}{pq}. \label{J_distr} \ee

The poles were shown in \cite{KHi} to be degenerate Killing horizons, which is in line with the general study of the DK solution made by Dietz and Hoenselaers \cite{DHo}. The area of each horizon is \cite{KHi}
\be
A_{\rm pole}=8\pi\kappa^2p^{-2}(1+p). \label{A_hor} \ee

The mass-quadrupole moment $Q$ of TS2 solution deserves a special mention as it has never been published with a correct sign for nearly four decades. The easiest way to calculate $Q$ is by using the expression of the corresponding Ernst potential, namely,
\bea
\E&=&(A-B)/(A+B), \nonumber\\
A&=&p^2(x^4-1)+q^2(y^4-1)-2ipqxy(x^2-y^2), \nonumber\\
B&=&2px(x^2-1)+2iqy(y^2-1). \label{E_pot} \eea
Then the procedure \cite{FHP} for the calculation of the Geroch-Hansen relativistic multipole moments \cite{Ger,Han} readily yields the desired formula for $Q$:
\be Q=-\frac{1}{4}M^3(1+3q^2),  \label{Q_TS2} \ee
and this $Q$ turns out to be {\it less} than the respective quadrupole moment of the Kerr solution. Consequently, the rectified $Q$ for all TS-solutions is
\be Q=-M^3\left(\frac{\delta^2-1}{3\delta^2}p^2+q^2\right).  \label{Q_TS} \ee

The geodesics in the TS2 spacetime were studied by various authors \cite{TSa2,GRu,BYo,Ern1}, and it was precisely the investigation of the equatorial geodesics that turned out crucial for demonstrating that the hypersurface  $x=1$ was not an event horizon \cite{GRu}. In the paper \cite{EEr} it was shown that the singularities of TS2 solution defined by the invariants of the Weyl conform tensor coincide with zeros of the denominator $A+B$ of the potential $\E$ from (\ref{E_pot}). Therefore, here we can omit the explicit form of these invariants calculated in \cite{EEr}. The function $A+B$ takes zero values at the poles, $x=1$, $y=\pm1$, and at one point of the equatorial ($y=0$) plane defined by the equation
\be p^2x^4+2px(x^2-1)-1=0, \label{x0_eq} \ee
whence for the real root $x=x_0>1$ we get the expression ($p>0$)
\be x_0=\frac{1}{2p}\left(\delta-1+\sqrt{3-\delta^2 -\frac{2}{\delta}(1-2p^2)}\right), \quad \delta\equiv\sqrt{1+(2p)^{2/3}(p^2-1)^{1/3}}. \label{x0} \ee

The point $(x_0,0)$ determines the massless ring singularity outside the symmetry axis. The ergoregion of TS2 solution is bounded by two SLS defined by the equation $f=0$, and the ring singularity lies on the inner surface if $p>0$, i.e., if the total mass is a positive quantity. The existence of the region with CTCs was first pointed out in \cite{GRu}, and this important region of causality violation is defined by the norm $\eta^\a\eta_\a$ of the axial Killing vector for which in \cite{KHi} a rather cumbersome expression was given. However, a simple formula for $\eta^\a\eta_\a$ was obtained long ago by Ernst \cite{Ern1}, and we write it down in the explicit form,
\bea
\eta^\a\eta_\a&=&\kappa^2(x^2-1)(1-y^2)f^{-1}-f\omega^2= \frac{\kappa^2(1-y^2)N}{p^2D}, \nonumber\\
N&=&p^2(x^2-1)[p^2(x^2-1)^2+4x(px^2+2x+p)+q^2(1-y^2)^2]^2 \nonumber\\
&&-4q^2(1-y^2)[p^2(x^2-1)(x^2-y^2)+2(px+1)(1-y^2)]^2, \label{norm} \eea
which is a good deal simpler than formula $(34d)$ of \cite{KHi}. Note that the function $D$ entering (\ref{norm}) is the denominator of the metric coefficient $f$ already defined in (\ref{mf_pol}).

In Fig.~1 we have plotted a typical shape of the ergoregion, region with CTCs and location of the ring singularity (the point at which the curves determining the outer boundary of the CTC region and inner boundary of the ergoregion touch each other) for a particular choice of the parameters $\kappa, p>0$.

\begin{figure}[htb]
\centerline{\epsfysize=90mm\epsffile{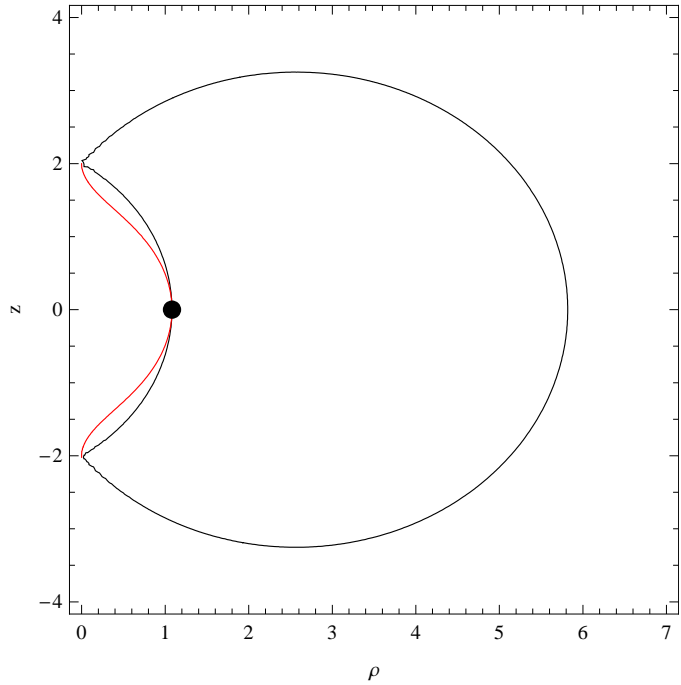}} \caption{The ergoregion (between two outer curves), the region with CTCs (between the axis and the inner curve), and the ring singularity in the TS2 solution (the particular choice of the parameters is $\k=2$, $p=3/5$, $q=4/5$).}
\end{figure}

Our idea to attribute the origin of the naked ring singularity and region with CTCs to the presence of a negative mass in the TS2 solution springs from the known analyses of the DK solution which provide a convincing evidence in favor of a direct connection between the negative mass of a rotating source and the corresponding ring singularity located on the SLS of that source \cite{Hoe,MRS,RMR}. Mention also that the recent paper \cite{GLS} supplies one with an additional indirect hint on the negative mass in TS2 solution since the only reasonable interpretation of the fact that the poles $x=1$, $y=\pm1$ have zero mass and non-zero angular momentum would be supposing that there are two sources associated with each pole which are endowed with equal positive and negative masses but unequal angular momenta. We are now turning to identifying analytically the negative mass distribution in the TS2 solution.

\section{The negative mass distribution in TS2 solution}

Since the metric function $\omega$ of the TS2 solution takes constant value on the interval $(-\kappa,\kappa)$ of the symmetry axis, the distributions of the Komar mass and Komar angular momentum \cite{Kom} on this rod can be analyzed with the aid of Tomimatsu's formulas \cite{Tom} (both $\omega$ and $\Omega$ to be evaluated at $x=1$, $y=z/\kappa$):
\be
M(z)=-\frac{1}{4}\omega[\Omega(z)-\Omega(-z)], \quad
J(z)=\frac{1}{4}\omega\{-2\kappa-\frac{1}{2}\omega[\Omega(z)-\Omega(-z)]\}, \label{MzJz} \ee
where $z$ stands for the interval $(-z,z)$, $|z|<\kappa$, of the symmetry axis containing the origin $z=0$, and $\Omega$ is the imaginary part of the complex potential $\E$ of the TS2 solution:
\be
\Omega=-\frac{4qyC}{D}, \quad C=p^2(x^2-1)^2(2x^2-y^2+1)+q^2(y^2-1)(y^4-1). \label{Om} \ee
Like in formulas (\ref{norm}), $D$ is the denominator of the function $f$.

On the interval $|z|<\kappa$ of the symmetry axis, $\omega$ and $\Omega$ take the form
\be
\omega=\frac{4\kappa(1+p)}{pq}, \quad \Omega=-\frac{4\kappa qz(z^2+\kappa^2)} {q^2(z^2+\kappa^2)^2+4\kappa^2(1+p)^2z^2}, \label{w_axis} \ee
so that for $M(z)$ and $J(z)$ we readily obtain
\be
M(z)=\frac{8\kappa^2(1+p)z(z^2+\kappa^2)} {p[q^2(z^2+\kappa^2)^2+4\kappa^2(1+p)^2z^2]}, \quad J(z)=\frac{2\kappa(1+p)}{pq}[M(z)-\kappa]. \label{Mz0} \ee

Restricting ourselves to the case of the positive total mass $M$, we must choose $p>0$ in the above formulas, and without lack of generality we can also put $q>0$ there as the negative values of $q$ only change sign in the expression of the total angular momentum $J$. Then for the point $z=z_{\rm max}$ at which $M(z)$ takes its maximum value we easily find from (\ref{Mz0})
\be
z_{\rm max}=\frac{\kappa} {\sqrt{1-p}}\left(\sqrt{1+p}-\sqrt{2p}\right), \label{zmax} \ee
so that the corresponding value $M(z_{\rm max})$ yields the amount of the positive mass $M_{\rm p}$ too, while the amount of the negative mass $M_{\rm n}$ can be found either directly from (\ref{Mz0}) and (\ref{zmax}) or simply by taking the difference between $M$ and $M_{\rm p}$ (of course, this is justified because $M(z)$ is monotonically increasing on the interval $0<z<z_{\rm max}$, and monotonically decreasing on the interval $z_{\rm max}<z<\kappa$). The resulting expressions for $M_{\rm p}$ and $M_{\rm n}$ are
\be
M_{\rm p}=\frac{2\kappa}{pq}, \quad M_{\rm n}=-\frac{2\kappa(1-q)}{pq}, \label{Mn} \ee
while the angular momentum $J_{\rm rod}$ of the rod is distributed between the momenta of the positive and negative masses in the following way:
\be
J_{\rm p}=\frac{2\kappa^2(2-pq)}{p^2(1-p)}, \quad J_{\rm n}=-\frac{4\kappa^2(1-q)}{p^2(1-p)}. \label{Jn} \ee

Therefore, we have established analytically the presence of the negative mass in TS2 solution and identified its location on the symmetry axis (the segments $\rho=0$, $z_{\rm max}<|z|<\kappa$). The ratio $|M_{\rm p}/M_{\rm n}|=1/(1-q)$ shows that the portion of the negative mass diminishes with increasing $q$, and tends to parity with the positive mass contribution in the limit $q\to0$.

\begin{figure}[htb]
\centerline{\epsfysize=80mm\epsffile{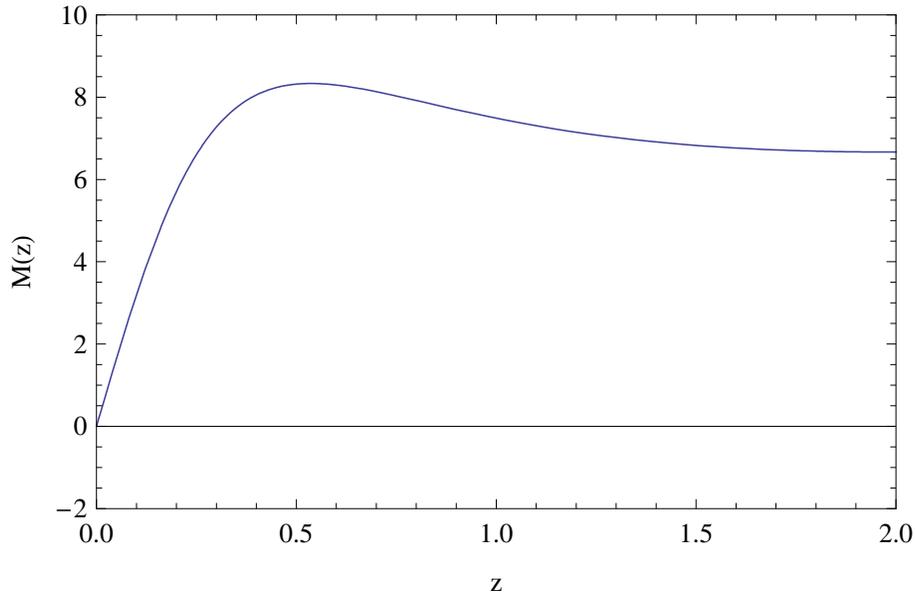}} \caption{The mass distribution on the segment $(-\kappa,\kappa)$ of the symmetry axis (the parameters are chosen as in Fig.~1).}
\end{figure}

\begin{figure}[htb]
\centerline{\epsfysize=80mm\epsffile{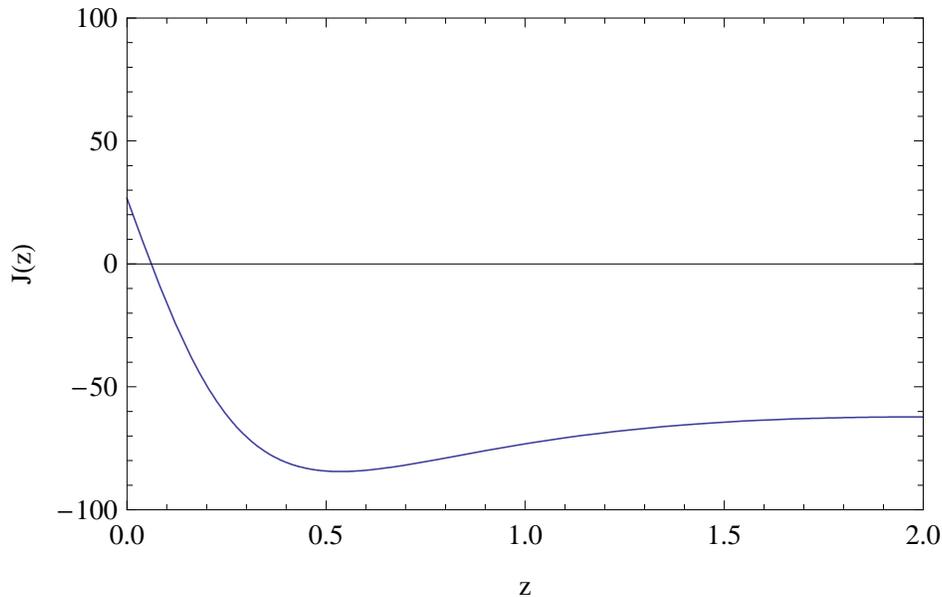}} \caption{The angular momentum distribution on the segment $(-\kappa,\kappa)$ of the symmetry axis (the parameters are chosen as in Fig.~1).}
\end{figure}

To illustrate the above general formulas, in Figs.~2 and 3 we have plotted the distributions $M(z)$ and $J(z)$ for a particular parameter choice $\k=2$, $p=3/5$, $q=4/5$ (the same as in Fig.~1), and one can see that on the interval $(-2(2-\sqrt{3}),2(2-\sqrt{3}))$ the mass is positive, taking its maximum value $M_{\rm p}=25/3$, while the parts $2(2-\sqrt{3})<|z|<2$ of the symmetry axis (the remaining parts of the rod) are characterized by a negative mass distribution with $M_{\rm n}=-5/3$, the overall total mass of the rod being $M=20/3$.

For the distribution of the angular momentum $J_{\rm rod}$, shown in Fig.~3, we have $J_{\rm p}=-200/9$, $J_{\rm n}=760/9$. To evaluate the total angular momentum, we also have to take into account the contribution of each pole $J_{\rm pole}=-40/3$; then
\be
J=J_{\rm p}+J_{\rm n}+2J_{\rm pole}=\frac{320}{9}, \label{Jt_ex} \ee
and it turns out that the contribution of the rotating negative mass into the angular momentum of the TS2 solution is the largest!

\section{A one-parameter generalization of the TS2 solution}

Now our interest consists in finding a generalization of the TS2 solution such that it would contain an additional parameter by varying which we could obtain evidence relating unambiguously the negative mass to the ring singularity. Mention that in the paper \cite{OSa} Oohara and Sato showed how the TS2 solution can be obtained from the DK solution of Kramer and Neugebauer \cite{KNe} in the special case of two equal corotating Kerr subextreme constituents. However, instead of considering the same equatorially symmetric binary configuration as in \cite{OSa} we shall employ a different approach requiring the knowledge of the axis value of the potential (\ref{E_pot}) that will give us some additional useful physical information.

We have found that on the upper part of the symmetry axis ($y=1$, $x=z/\kappa$) the potential $\E$ of the TS2 solution (\ref{E_pot}) can be cast into the form ($q>0$)
\be
e(z)=\frac{z-\frac{\kappa}{p}(1-\sqrt{q})(1-i\sqrt{q})} {z+\frac{\kappa}{p}(1-\sqrt{q})(1+i\sqrt{q})} \cdot\frac{z-\frac{\kappa}{p}(1+\sqrt{q})(1+i\sqrt{q})} {z+\frac{\kappa}{p}(1+\sqrt{q})(1-i\sqrt{q})}. \label{TS_axis} \ee
This means, taking into account that the axis value of a single Kerr solution is given by the expression
\be
e(z)=\frac{z-m-ia} {z+m-ia}, \label{Kerr_axis} \ee
where $m$ is the mass parameter and $a$ the rotational parameter, that the TS2 solution can be {\it formally} considered as a non-linear superposition of two non-identical Kerr black holes -- one with the mass $\frac{\kappa}{p}(1-\sqrt{q})$ and angular momentum per unit mass $-\frac{\kappa}{p}\sqrt{q}(1-\sqrt{q})$, and the other with the mass $\frac{\kappa}{p}(1+\sqrt{q})$ and angular momentum per unit mass $\frac{\kappa}{p}\sqrt{q}(1+\sqrt{q})$ -- both black holes placed at the origin of coordinates and counterrotating. Since the terms $(1\pm\sqrt{q})$ are positive, then for $p>0$ the masses are positive too. Hence the question arises: where does the negative mass, the presence of which in the TS2 solution has been established earlier, come from?

Probably an immediate simple answer to this question could be the following: a negative mass is needed to prevent two overlapping black holes from forming a unique Kerr black hole. On the other hand, it should be also remembered that we have a non-linear superposition in the presence of a strong gravitational field that can drastically change the initial interpretation of a binary configuration. To clarify the situation, we find it instructive to consider an immediate generalization of the axis data (\ref{TS_axis}) of TS2 solution by introducing shifts of the black-hole constituents along the symmetry axis:
\be
e(z)=\frac{z-z_0-\frac{\kappa}{p}(1-\sqrt{q})(1-i\sqrt{q})} {z-z_0+\frac{\kappa}{p}(1-\sqrt{q})(1+i\sqrt{q})} \cdot\frac{z+z_0-\frac{\kappa}{p}(1+\sqrt{q})(1+i\sqrt{q})} {z+z_0+\frac{\kappa}{p}(1+\sqrt{q})(1-i\sqrt{q})}, \label{TSg_axis} \ee
where the first constituent is shifted from the origin at $z_0$, and the second at $-z_0$. Now we have a more general system of two non-identical separated Kerr black holes which in the limit $z_0\to 0$ leads to the TS2 configuration. A remarkable property of this system is that it has the same total mass and total angular momentum as the TS2 solution, but its quadrupole moment generalizes formula (\ref{Q_TS2}):
\be Q=-\frac{1}{4}M^3(1+3q^2)+Mz_0^2.  \label{Q_TS2g} \ee

If we succeed to show that the solution defined by (\ref{TSg_axis}) contains an intermediate region with negative mass, then we may consider the TS2 spacetime as a final state of such special two-body system, with the negative mass redistributed within the rod and the poles in the limit $z_0=0$. So what we have to do now is to construct an exact solution corresponding to the axis data (\ref{TSg_axis}) and vary in it the parameter $z_0$ starting from some sufficiently large value and then go decreasing $z_0$ and calculating the Komar masses and angular momenta of the constituents and of the intermediate region which separates them. As will be seen later, the lower constituent during such evolution will always remain subextreme, while the upper constituent is first subextreme, then becomes a hyperextreme object and eventually ends up as an extreme source at $z_0=0$. Therefore, we have to make use of the analytically extended double-Kerr (AEDK) solution  \cite{MRS,MRu} obtained with the aid of Sibgatullin's integral method \cite{Sib,MSi} in order to be able to describe all possible changes of the binary system under consideration (the original DK solution is only valid for the description of the subextreme Kerr constituents). We recall that the complex potential $\E$ of the AEDK solution has the form
\bea
{\cal E}&=&E_+/E_-, \nonumber\\
E_\pm&=&-(\a_1-\a_2)(\a_3-\a_4)(X_1X_2r_1r_2+X_3X_4r_3r_4)+(\a_1-\a_3)(\a_2-\a_4) \nonumber\\ &&\times(X_1X_3r_1r_3+X_2X_4r_2r_4) -(\a_1-\a_4)(\a_2-\a_3)(X_1X_4r_1r_4+X_2X_3r_2r_3), \nonumber\\
&&\pm(\a_3-\a_4)[(\a_1-\a_3)(\a_1-\a_4)X_2r_2 -(\a_2-\a_3)(\a_2-\a_4)X_1r_1] \nonumber\\ &&\pm(\a_1-\a_2)[(\a_1-\a_3)(\a_2-\a_3)X_4r_4 -(\a_1-\a_4)(\a_2-\a_4)X_3r_3], \nonumber\\  X_i&=&\frac{(\a_i-\bar\beta_1)(\a_i-\bar\beta_2)}{(\a_i-\beta_1)(\a_i-\beta_2)}, \quad r_i=\sqrt{\rho^2+(z-\a_i)^2},  \label{DK_an} \eea
where $\beta_l$ are arbitrary complex parameters, and $\a_n$ can take arbitrary real values or occur in complex conjugate pairs. Since $\beta_l$ are simple poles in the axis data, then for our particular case we find from (\ref{TSg_axis})
\be
\beta_1=z_0-\frac{\kappa}{p}(1-\sqrt{q})(1+i\sqrt{q}), \quad \beta_2=-z_0-\frac{\kappa}{p}(1+\sqrt{q})(1-i\sqrt{q}). \label{betas} \ee
On the other hand, the form of $\a_n$ corresponding to (\ref{TSg_axis}) can be obtained by solving (for $z$) the quartic equation
\be
(1+q)[(z^2-\kappa^2-z_0^2)^2-4\kappa^2z_0^2]+8\kappa^2z_0\sqrt{q}z=0, \label{eq_alg} \ee
which arises from the condition
\be
e(z)+\bar e(z)=0 \label{cond_alg} \ee
of Sibgatullin's methos.

Although Eq.~(\ref{eq_alg}) can be solved analytically, the resulting expressions are not wieldy to work with. Because we are basically interested in the qualitative picture of the evolution of the system, in what follows we shall assign some particular values to the parameters $p$, $q$ and $\kappa$, and consider different values of $z_0$ for which we shall be finding the corresponding concrete $\a_n$ by solving numerically Eq.~(\ref{eq_alg}). Once the values of $\beta_l$ and $\a_n$ are known, the quantities $X_n$ in (\ref{DK_an}) are also known, so that the corresponding metric functions and Komar characteristics of the binary system can be worked out with the aid of formulas (\ref{mf_DK})-(\ref{MJ_gen}) of Appendix.

In table 1 we have described four particular configurations for the relatively large values of $z_0$ (the parameters $p$, $q$ and $\kappa$ are fixed at 3/5, 4/5 and 1, respectively), where $M_1$, $M_2$ and $M_0$ denote the Komar masses of the upper constituent, lower constituent and intermediate region, respectively; $J_1$, $J_2$ and $J_0$ are the corresponding angular momenta. One can see that at first the two constituents are subextreme sources (all $\a_n$ are real) with positive masses, then, at smaller separations, the upper constituent becomes a hyperextreme object ($\a_2=\bar\a_1$). The intermediate region has a small negative mass $M_0$ introduced by the NUT parameters of the sources. No ring singularity appears in those configurations because the sources have positive masses.

\begin{table}[htb]
\caption{Evolution of the generalized TS2 solution with varying $z_0$ ($z_0>\kappa$). The parameters $p$, $q$ and $\kappa$ are assigned the values $0.6$, $0.8$ and $1$, respectively.}
\begin{center}
\begin{tabular}{lccccccc}
\hline $z_0$ & $8$ & $6$ &
$4$ & $2$ \\ \hline $M_0$ & $-0.0015$ & $-0.0002$ & $-0.014$ & $-0.251$ \\ $J_0$ & $-0.258$ & $-0.385$ &
$-0.719$ & $-2.824$ \\
$M_1$ & $0.168$ & $0.157$ & $0.139$ & $0.106$ \\ $J_1$ & $-0.039$ & $-0.042$ &
$-0.050$ & $-0.112$ \\ $M_2$ & $3.167$ & $3.176$ & $3.208$ & $3.479$ \\ $J_2$ & $9.186$ & $9.316$ &
$9.658$ & $11.825$ \\ $\beta_1$ & $7.824-0.157i$ & $5.824-0.157i$ & $3.824-0.157i$ & $1.824-0.157i$ \\ $\beta_2$ & $-11.157+2.824i$ & $-9.157+2.824i$ &
$-7.157+2.824i$ & $-5.157+2.824i$ \\ $\a_1$ & $8.048$ & $6.001-0.028i$ & $4.002-0.099i$ & $2.02-0.251i$ \\ $\a_2$ & $7.952$ & $6.001+0.028i$ & $4.002+0.099i$ & $2.02+0.251i$ \\ $\a_3$ & $-6.59$ & $-4.591$ & $-2.596$ & $-0.639$ \\ $\a_4$ & $-9.411$ & $-7.41$ &
$-5.408$ & $-3.4$ \\ \hline
\end{tabular}
\end{center}
\end{table}

In table 2 the configurations undergo changes at smaller separations due to the spin-spin interaction of the three regions. The positive and negative masses of the lower constituent and intermediate region grow in value and then interchange their signs: at $z_0=0.8$, $M_2$ is already negative, while $M_0$ is positive. The ring singularity appears precisely after $M_2$ becomes negative and is located on the SLS of the lower constituent.

\begin{table}[htb]
\caption{Evolution of the generalized TS2 solution with varying $z_0$ ($z_0<\kappa$). The values of $p$, $q$, $\kappa$ are the same as in table~1.}
\begin{center}
\begin{tabular}{lccccccc}
\hline $z_0$ & $0.9$ & $0.8$ &
$0.5$ & $0.001$ \\ \hline $M_0$ & $-30.18$ & $18.115$ & $4.693$ & $3.334$ \\ $J_0$ & $-193.434$ & $112.614$ &
$26.818$ & $15.664$ \\
$M_1$ & $0.084$ & $0.078$ & $0.055$ & $0.00016$ \\ $J_1$ & $-0.966$ & $-1.194$ &
$-1.974$ & $-3.331$ \\ $M_2$ & $33.430$ & $-14.86$ & $-1.414$ & $-0.001$ \\ $J_2$ & $203.288$ & $-102.53$ &
$-15.955$ & $-3.444$ \\  $\beta_1$ & $0.724-0.157i$ & $0.624-0.157i$ & $0.324-0.157i$ & $-0.175-0.157i$ \\ $\beta_2$ & $-4.057+2.824i$ & $-3.957+2.824i$ &
$-3.657+2.824i$ & $-3.158+2.824i$ \\ $\a_1$ & $1.143-0.529i$ & $1.105-0.549i$ & $1.035-0.548i$ & $1-0.032i$ \\ $\a_2$ & $1.143+0.529i$ & $1.105+0.549i$ & $1.035+0.548i$ & $1+0.032i$ \\ $\a_3$ & $-0.01$ & $-0.039$ & $-0.222$ & $-0.968$ \\ $\a_4$ & $-2.277$ & $-2.172$ &
$-1.847$ & $-1.032$ \\ \hline
\end{tabular}
\end{center}
\end{table}

In Fig.~4 we illustrate the appearance and evolution of the ring singularity on SLS of the constituent with negative mass for different small $z_0$, which leaves no doubt that the origin of this singularity is the negative mass.

\begin{figure}[htb]
\centerline{\epsfysize=120mm\epsffile{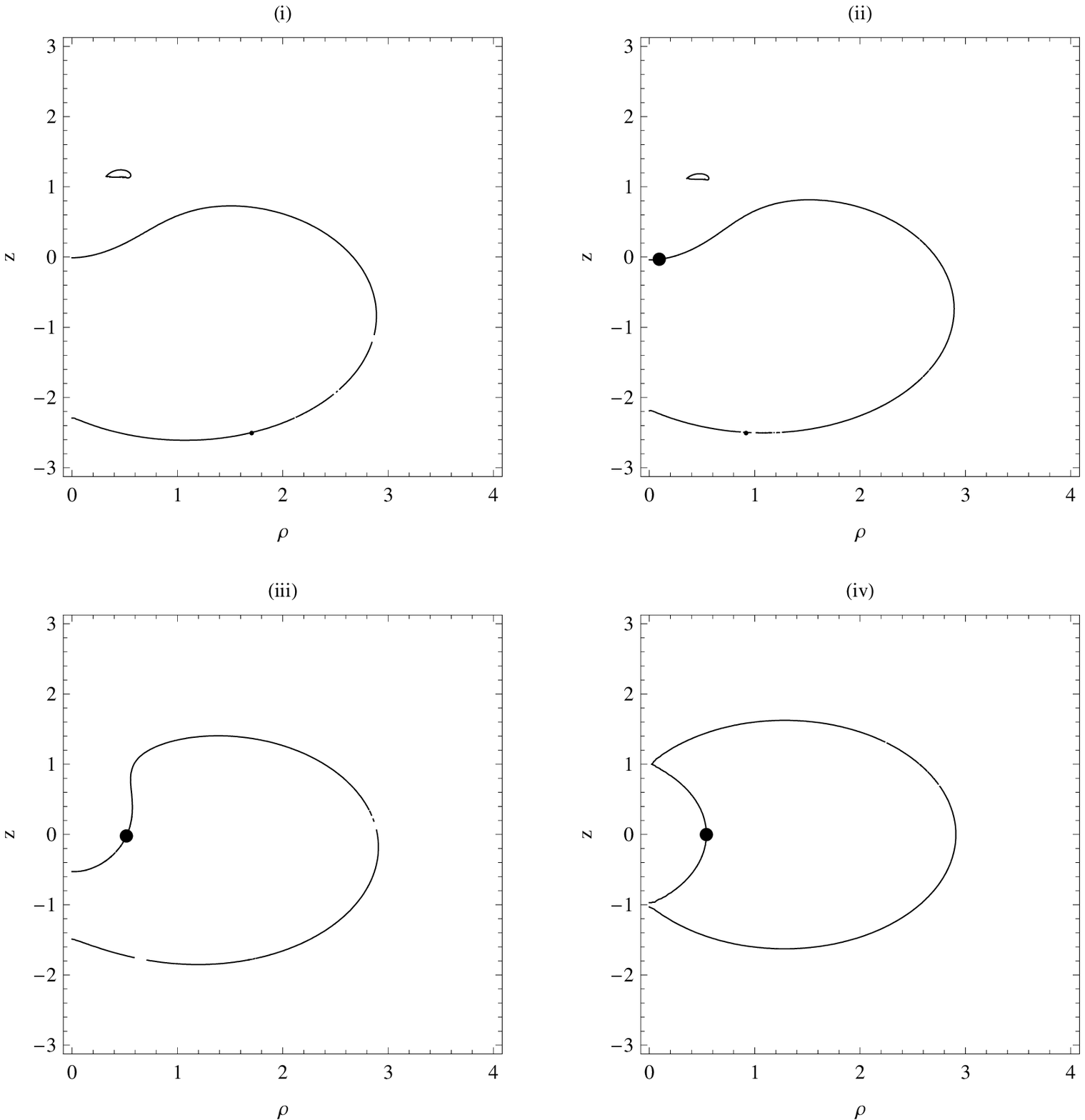}} \caption{The SLS and ring singularities for particular configurations defined by $z_0=0.9, 0.8, 0.2, 0.001$. The ring singularity is absent in Fig.~5(i) because the mass of the lower constituent is still positive.}
\end{figure}

It should be especially remarked that the appearance of the region with CTCs in the configurations described by the above generalization of TS2 solution coincides with the appearance of the ring singularity and is also linked to the SLS of the lower constituent possessing a negative mass. We have checked numerically that the ring singularity always lies on the boundary of the CTC region, the latter boundary and the SLS of the lower constituent touching each other at the point determining the ring singularity. However, we have been unable to get high-quality plots of the particular CTC regions because of the corresponding complicated analytical expression of the norm of the axial Killing vector which was causing problems of the numerical computer processing.

\section{The TS2 solution with negative total mass}

Once we have established that the TS2 solution with $M>0$ ($p>0$) describes distributions of positive and negative Komar masses in which the positive mass prevails over the negative one, it seems quite logic to briefly discuss the opposite case when the negative mass prevails over the positive one and hence the total mass $M$ becomes a negative quantity. It is interesting in particular to find out whether in the new situation we are going to have the same picture for the region with CTCs and the same location of the ring singularity as in Fig.~1. Like the previous case of positive $M$, the case $M<0$ ($p<0$) can be readily treated with the aid of formulas (\ref{mf_pol}), (\ref{pol4}), (\ref{E_pot}), (\ref{x0_eq}) and (\ref{norm}).

\begin{figure}[htb]
\centerline{\epsfysize=80mm\epsffile{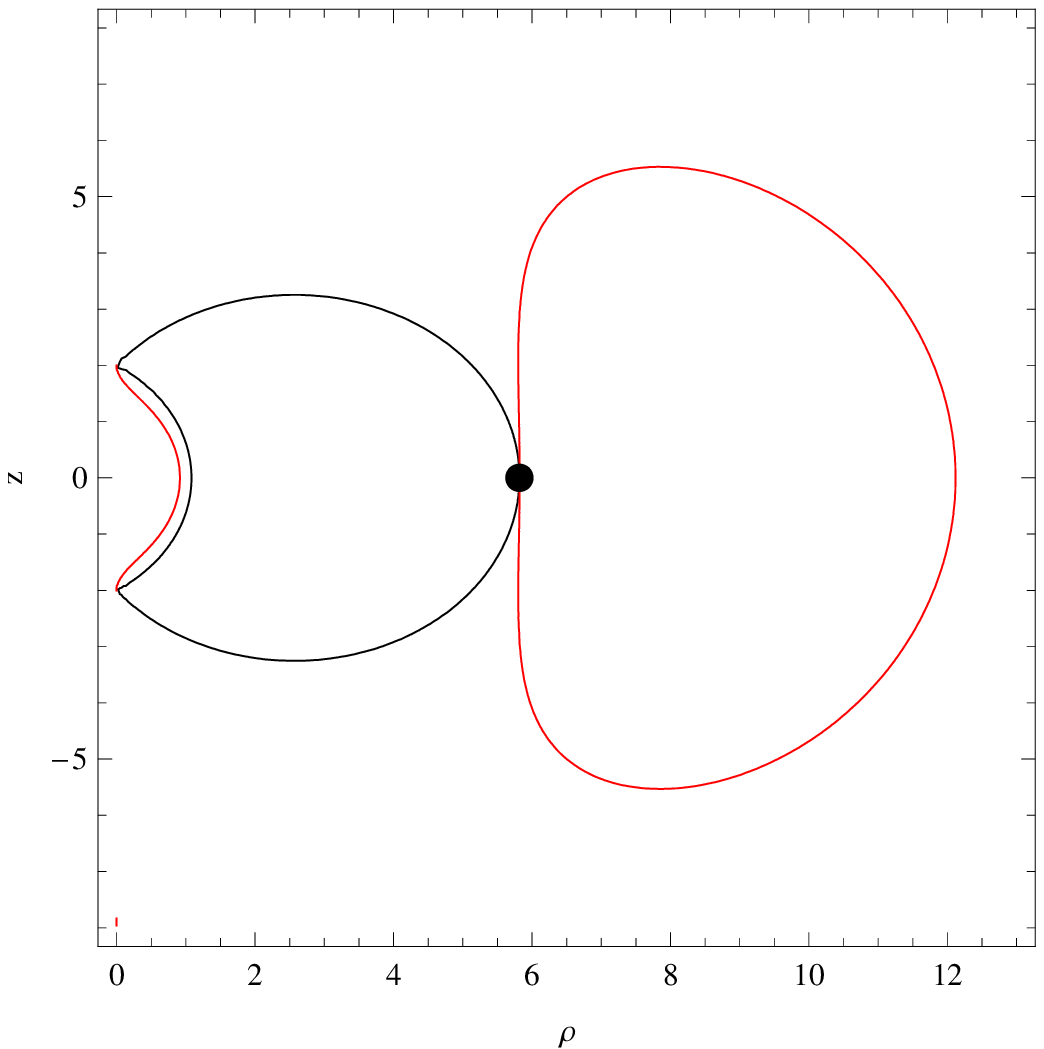}} \caption{The regions with CTCs and the ring singularity in the TS2 solution with negative total mass.}
\end{figure}

In Fig.~5 we have plotted the SLS, the region of CTCs and the ring singularity of the TS2 solution for a particular parameter choice $p=-3/5$, $q=4/5$, $\kappa=2$ which provides us with the same absolute value of mass $M$ and the same angular momentum $J$ as in the example used for plotting Fig.~1. It turns out that although the SLS in both figures are identical, the figures differ considerably in the shapes of the CTC regions and positions of the ring singularities. The first important distinction is the apearance in Fig.~5 of the second region with CTCs which has a toroidal topology, and the second distinction is that the ring singularity in the case $M<0$ appears on the outer boundary of the ergoregion, thus reminding the position of ring singularities in the equilibrium configurations involving constituents with negative masses in the AEDK solution \cite{MRS}. Observing that the first CTC region (having topology of a sphere) in Fig.~5 does not touch the inner SLS outside the axis and does not contain any ring singularity, we arrive at a generic conclusion that the massless ring singularity can be regarded as a direct outcome of the SLS touching the boundary of the CTC region, and all this becomes possible only due to presence of a negative mass. Analytically, the singularity is defined in the prolate spheroidal coordinates by the point $(x_0,0)$, $x_0$ having the form
\be x_0= \left\{\begin{array}{rl} -\dfrac{1}{2p}\left(\delta+1+\sqrt{3-\delta^2 +\frac{2}{\delta}(1-2p^2)}\right),& \quad -\dfrac{1}{\sqrt{2}}<p<0 \vspace{0.2cm}\\ \dfrac{1}{2p}\left(\delta-1-\sqrt{3-\delta^2 -\frac{2}{\delta}(1-2p^2)}\right),& \quad -1<p<-\dfrac{1}{\sqrt{2}}\\ \end{array}\right. \label{x0_neg} \ee
where $\delta$ is the same as in (\ref{x0}).

It looks interesting to compare Fig.~5 with an analogous figure in the case of a single Kerr solution with negative mass. An extra interest to do this consists in a necessity to correct a long-standing wrong belief that the Boyer-Lindquist coordinates~\cite{BLi} commonly used for the presentation and analysis of the Kerr metric are equally applicable to the situations with positive and negative masses. Then let us write down the metric functions $f$, $\gamma$ and $\omega$ of the Kerr solution in the subextreme case using the Ernst representation in prolate spheroidal coordinates:
\be
f=\frac{p^2x^2+q^2y^2-1}{(px+1)^2+q^2y^2}, \quad e^{2\gamma}=\frac{p^2x^2+q^2y^2-1}{p^2(x^2-y^2)}, \quad \omega=-\frac{2\kappa q(1-y^2)(px+1)}{p(p^2x^2+q^2y^2-1)}, \label{f_ker} \ee
the corresponding total mass and total angular momentum having the form
\be
M=\frac{\kappa}{p}, \quad J=\frac{\kappa^2q}{p^2}, \quad p^2+q^2=1. \label{M_ker} \ee

Like in the case of the TS2 solution, the negative values of $M$ are defined by $p<0$, and the Kerr solution for $p<0$ has a singularity at the point
\be
x_0=-\frac{1}{p}, \quad y_0=0 \quad \Longleftrightarrow \quad \rho_0=-\frac{|q|}{p}\k=\left|\frac{J}{M}\right|, \quad z_0=0, \label{x_ker} \ee
as can be easily seen by considering the denominator of the function $f$ in (\ref{f_ker}). The singularity lies off the symmetry axis (since $x_0>1$ or $\rho_0>0$) in the equatorial plane and, therefore, is a naked ring singularity. It is also a curvature singularity, as follows from the form of the corresponding scalar invariant $I_1$ of the Weyl tensor,
\be
I_1=-\frac{p^2}{2\k^2(px+1-iqy)^3}, \label{I1} \ee
which can be calculated with the aid of the Economou-Ernst procedure \cite{EEr} (the second invariant $I_2$ is equal to zero identically).

The SLSs of the Kerr solution with positive and negative masses are identical, being defined by the equation
\be
p^2x^2+q^2y^2-1=0, \label{SLS} \ee
and since the point $x=-1/p$, $y=0$ fulfils (\ref{SLS}), the ring singularity is located on the boundary of the ergoregion.

On the other hand, the ring singularity also belongs to the boundary of the region with CTCs. Indeed, the norm of the axial Killing vector of the Kerr solution has the form
\bea
\eta^\a\eta_\a&=&\frac{\kappa^2(1-y^2)N}{p^2[(px+1)^2+q^2y^2]}, \nonumber\\ N&=&[(px+1)^2+q^2]^2-p^2q^2(x^2-1)(1-y^2), \label{norm_K} \eea
and by a direct check one verifies that $x=-1/p$, $y=0$ satisfies $N=0$ identically, while the CTC region itself is defined by the inequality $N<0$. Hence, the ring singularity is a locus where the SLS and boundary of the CTC region touch each other. In the equatorial plane ($y=0$) the region with CTCs extends from $x=x_0$ to $x=x_1$ where
\be x_1=-\frac{1}{p}\left(1-\Delta+\frac{q^2}{3\Delta}\right), \quad \Delta\equiv\left(q^2\sqrt{1+\frac{1}{27}q^2}\right)^{1/3}, \label{x1} \ee
and it can be readily verified that $x_1>x_0$ and the norm $\eta^\a\eta_\a$ is negative definite on the interval  $x_0<x<x_1$, $y=0$.

The ring singularity of the Kerr solution with negative mass is massless, and all the mass is contained in the central rod $x=1$ of the symmetry axis, as can be readily seen with the aid of formulas (\ref{MzJz}) taking into account that the function $\Omega$ of the Kerr solution is given by the expression
\be \Omega=-\frac{2qy}{(px+1)^2+q^2y^2}. \label{Om_ker} \ee

\begin{figure}[htb]
\centerline{\epsfysize=80mm\epsffile{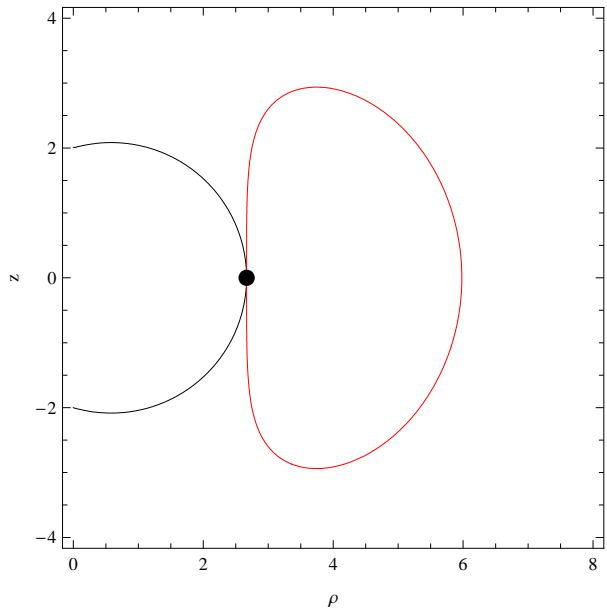}} \caption{The SLS (black curve), region with CTCs (inside the red curve) and ring singularity in the Kerr solution with negative mass.}
\end{figure}

In Fig.~6 we have plotted the SLS, the region with CTCs and the massless ring singularity (the point $\rho_0=8/3,z_0=0$) of the Kerr solution by choosing the same particular values of parameters $p=-3/5$, $q=4/5$, $\kappa=2$ as in the case of Fig.~5. Obviously, both figures 5 and 6 are quite similar qualitatively.

Now, what happens if instead of analyzing the Kerr solution endowed with negative mass directly in the coordinates $(x,y)$, one first passes to the standard Boyer-Lindquist coordinates $(r,\vartheta)$ via the formulas
\be
x=\frac{r-M}{\kappa}, \quad y=\cos\vartheta, \label{BL} \ee
or simply starts with a representation of the Kerr metric in these coordinates? It is easy to see that while the choice $M>0$ yields an extension of the Kerr metric, the case $M<0$ leads to a contraction of the negative-mass Kerr spacetime because $r=\cos\vartheta=0$ then corresponds to location of a massless ring singularity which is a point off the symmetry axis! In this respect it is worth noting that Boyer and Lindquist in their paper \cite{BLi} explicitly state that their consideration is restricted to the case $M>0$, and in no place they mention that their coordinates are also a good option for $M<0$. In any case, a reasonable coordinate change for $M<0$ might be (\ref{BL}) with $M$ substituted by $|M|$, or by any other positive quantity which would have some physics behind it. Our own opinion is that the use of the spheroidal coordinates alone is sufficient for demonstrating the non-physical character of the Kerr solution with negative mass, especially taking into account that the singularity in this case is located outside the symmetry axis.

The above said about the subextreme Kerr solution with negative mass can be summarized in the form of the following theorem.

\medskip

\noindent{\bf Theorem.} {\it The subextreme Kerr solution with negative mass develops a massless naked ring singularity off the symmetry axis defined by the point $x=1/p$, $y=0$ ($\rho=|J/M|$, $z=0$ in cylindrical coordinates) in the equatorial plane. The singularity is a locus where the SLS touches the region with CTCs, and the Kerr metric in this case does not describe a black hole.}

\section{Discussion and conclusions}

The analysis of the TS2 solution carried out in the present paper suggests that the physical interpretation proposed for this solution long ago by Gibbons and Russell-Clark as describing the exterior field of a specific  star whose prolateness is larger than that of a Kerr black hole still remains the most reasonable one. Indeed, the presence of a massless ring singularity, region with CTCs and negative mass distribution seems to discard the TS2 metric as a relevant candidate able to describe the final stage of the gravitational collapse; on the other hand, it is not excluded that a physically meaningful interior solution might exist which could be matched with the TS2 spacetime on an appropriate boundary outside the problematic regions.

While the analytical formulas obtained by us give a clear picture of how the negative mass is distributed within the central rod $x=1$, $y^2<1$ of the TS2 solution, it is still not clear at the moment how to estimate the amount of negative mass contained in the poles $x=1$, $y=\pm1$ except stating that it is equal in absolute value to the amount of positive mass. At the same time, it is not difficult to answer a question about where the negative mass comes from. The DK metric is a non-linear superposition of two Kerr-NUT solutions \cite{NTU,DNe}, and the negative mass is introduced by the semi-infinite NUT sources, as it follows directly from the study of the latter sources either in the case of the pure NUT solution \cite{MRu1} or in the case of the combined Kerr-NUT spacetime \cite{MMR}.
The two NUT sources in the DK solution can be eliminated exclusively through imposing and solving the condition of asymptotic flatness and the axis condition. However, in the case of the TS2 solution only the former condition is satisfied, which means that some specific non-zero NUT sources are still present in this solution, giving rise to the negative mass. The TS2 axis data (\ref{TS_axis}) well illustrates a simple idea that a product of two single Kerr data does not necessarily lead to a binary black-hole configuration, but rather to a specific system of two Kerr-NUT sources.

We hope that we have been able to produce enough evidence in favor of an intrinsic connection existing between the negative mass, region with CTCs and massless ring singularity. The consideration of the TS2 and Kerr metrics with negative mass was crucial for establishing that it is precisely the `contact' of the ergosphere and boundary of the region with CTCs that determines location of the naked ring singularity in the vacuum case. The fact that a correct interpretation of the negative-mass Kerr spacetime has become possible only after a good understanding of a massless ring singularity of the TS2 metric demonstrates once again that the latter metric is a remarkable tool for testing new ideas in the field of exact solutions.

Finally, we would like to remark that our discussion of the negative-mass Kerr solution casts doubts on the credibility of the claim made in the paper \cite{APe} about the existence of a negative-mass regular black hole surrounded by a massive fluid ring. Since the claim was made on the basis of a purely numerical approach to the problem and hence cannot be independently checked analytically, we restrict ourselves to only pointing out possible sources of error: ($i$) the use of the coordinates that do not allow to identify the presence of a ring singularity, ($ii$) failure to reproduce unphysical regions by the numerical methods of paper \cite{APe}, and ($iii$) automatic attributing the properties of event horizons to null hypersurfaces which can be not true \cite{GKo}. In any case, a black hole with negative Komar mass, i.e., an object exhibiting repulsive gravity, looks to contradict the very essence of the notion `black holes'.

\section*{Acknowledgements}
The author wishes to dedicate this paper to Humitaka Sato and Akira Tomimatsu on the occasion of forthcoming 40th anniversary of the first publication of their solution. He is also grateful to the Relativity and Cosmology group at the School of Mathematical Sciences, Queen Mary, London, especially to Malcolm MacCallum and Juan Valiente Kroon, for hospitality during his short visits in 2011 when part of this work was written.

\appendix
\section{Metric functions of the analytically extended DK solution} 
%

The AEDK solution is the $N=2$ specialization of the extended multi-soliton vacuum metric considered in the paper \cite{MRu}. Its functions $f$, $\gamma$ and $\omega$ entering the line element
\be
d s^2=f^{-1}[e^{2\gamma}(d\rho^2+d z^2)+\rho^2d\varphi^2]-f(d
t-\omega d\varphi)^2, \label{papa2}
\ee
are defined by the following expressions:
\bea f&=&\frac{E_+\bar E_-+\bar
E_+E_-}{ 2E_-\bar E_-}, \quad e^{2\gamma}=\frac{E_+\bar
E_-+\bar E_+E_-}{2\lambda\bar\lambda\prod_{n=1}^{4}r_n}, \quad
\omega=\frac{2i(G\bar E_--\bar GE_-)}{E_+\bar E_-+\bar
E_+E_-},\nonumber\\
G&=&-\s\Lambda+(z+\bar\s)\Gamma+\sum_{1\leq i<j\leq4}
\lambda_{ij}(\a_i+\a_j)r_ir_j -\sum_{i=1}^4\gamma_i(\a_{i'}+\a_{j'}+\a_{k'})r_i, \quad \nonumber\\
&&(i',j',k'\neq i;\,\, i'<j'<k') \nonumber\\ \Lambda&=&\sum_{1\leq i<j\leq4}\lambda_{ij}r_ir_j, \quad
\Gamma=\sum_{i=1}^4\gamma_ir_i,\nonumber\\
\lambda_{ij}&=&(-1)^{i+j}(\a_i-\a_j)(\a_{i'}-\a_{j'})X_iX_j,
\quad (i',j'\neq i,j;\,\, i'<j') \nonumber\\
\gamma_{i}&=&(-1)^{i}(\a_{i'}-\a_{j'})(\a_{i'}-\a_{k'})
(\a_{j'}-\a_{k'})X_i, \quad (i',j',k'\neq i;\,\, i'<j'<k')
\nonumber\\ \s&:=&\beta_1+\beta_2,
\quad \lambda:=\sum_{1\leq i<j\leq4}\lambda_{ij}. \label{mf_DK} \eea
The explicit form of $E_\pm$, $X_i$ and $r_i$ is given in (\ref{DK_an}).

The Komar mass and angular momentum of the subextreme constituent can be calculated with the aid of Tomimatsu's formulas \cite{Tom}
\be M_a=-\frac{1}{8\pi}\int_{H_a}
\omega\Omega_{,z}d\varphi d z, \quad
J_a=\frac{1}{8\pi}\int_{H_a}\omega
\left(-1-\frac12\omega\Omega_{,z}\right)d\varphi d z, \label{MJ_Tom} \ee
while for the calculation of the Komar characteristics of the hyperextreme object the more general formulas must be used \cite{MRS}:
\bea M_a&=&\frac{1}{4}\Bigl(\int_{z_l}^{z_u}[\rho(\ln f)_{,\rho}-
\omega\Omega_{,z}]_{\rho=\rho_0}d z +\int_{0}^{\rho_0}[\rho(\ln f)_{,z}+
\omega\Omega_{,\rho}]_{z=z_u}d\rho \nonumber\\ &&-\int_{0}^{\rho_0}[\rho(\ln
f)_{,z}+ \omega\Omega_{,\rho}]_{z=z_l}d\rho\Bigr), \nonumber\\
J_a&=&-\frac{1}{8}\Bigl(\int_{z_l}^{z_u} [2\omega-2\rho\omega(\ln
f)_{,\rho}+ (\rho^2f^{-2}+\omega^2)\Omega_{,z}]_{\rho=\rho_0}d z
\nonumber\\ &&-\int_{0}^{\rho_0}[2\rho\omega(\ln f)_{,z}+
(\rho^2f^{-2}+\omega^2)\Omega_{,\rho}]_{z=z_u}d\rho \nonumber\\
&&+\int_{0}^{\rho_0}[2\rho\omega(\ln f)_{,z}+
(\rho^2f^{-2}+\omega^2)\Omega_{,\rho}]_{z=z_l}d\rho\Bigr), \label{MJ_gen} \eea
where $\Omega$ denotes the imaginary part of the Ernst potential $\E$. For other notations we refer to papers\cite{Tom,MRS}.

\end{document}